\documentclass{aa}

\input{psfig.sty}
\usepackage{graphicx}
\usepackage{natbib}

\def\beq{\begin{equation}}
\def\enq{\end{equation}}

\title{INTEGRAL/IBIS observations of a hard X-ray outburst in high mass X-ray binary 4U 2206+54}

\institute{National Astronomical Observatories, Chinese Academy of
Sciences, Beijing 100012, China}

\author{Wei Wang}

\offprints{W. Wang, wangwei@bao.ac.cn}
\date{Received }

\authorrunning{W. Wang}
\titlerunning{A hard X-ray outburst in 4U 2206+54}

\begin{document}

\abstract {}{4U 2206+54 is a wind-fed high mass X-ray binary with
a main-sequence donor star. The nature of its compact object was
recently identified as a slow-pulsation magnetized neutron star. }
{INTEGRAL/IBIS observations have a long-term hard X-ray monitoring
of 4U 2206+54 and detected a hard X-ray outburst around 15
December 2005 combined with the RXTE/ASM data. }  {The hard X-ray
outburst had a double-flare feature with a duration of $\sim$ 2
days. The first flare showed a fast rise and long time decaying
light curve about 15 hours with a peak luminosity of $\sim 4\times
10^{36}$ erg s$^{-1}$ from 1.5 -- 12 keV and a hard spectrum (only
significantly seen above 5 keV). The second one had the mean hard
X-ray luminosity of $1.3\times 10^{36}$ erg s$^{-1}$ from 20 --
150 keV with a modulation period at $\sim 5550$ s which is the
pulse period of the neutron star in 4U 2206+54; its hard X-ray
spectrum from 20 -- 300 keV can be fitted with a broken power-law
model with the photon indexes $\Gamma_1 \sim 2.3,\ \Gamma_2 \sim
3.3$, and the break energy is $E_b \sim 31$ keV or a
bremsstrahlung model of $kT\sim 23$ keV.} {We suggest that the
hard X-ray flare could be induced by suddenly enhanced accretion
dense materials from stellar winds hitting the polar cap region of
the neutron star. This hard X-ray outburst may be a link to
supergiant fast X-ray transients though 4U 2206+54 has a different
type of companion.}


\maketitle

\section{Introduction}

Massive X-ray binary 4U 2206+54 was discovered by the Uhuru
satellite (Giacconi et al. 1972). The optical counterpart was
identified as an O9.5V star with a high He abundance and has a
distance of $\sim 2.6$ kpc (Blay et al. 2006). Without a
circumstellar disk around the donor, the material for accretion
and production of high energy emission must come from stellar wind
(Negueruela \& Reig 2001). The wind terminal velocity of 4U2206+54
has a low value of $\sim 350$ km s$^{-1}$ (Ribo et al. 2006), so
assuming an eccentric orbit and using the Bondi-Hoyle formalism
wind-fed accretion could produce X-ray luminosity and variability
($L_x\sim 10^{33}-10^{35}$ erg s$^{-1}$) as seen by RXTE,
BepposSAX, Swift and INTEGRAL (Torrejon et al. 2004; Masetti et
al. 2004; Blay et al. 2005; Corbet et al. 2007; Wang 2009).

X-ray monitoring of 4U 2206+54 by RXTE suggested a modulation
period of 9.6 days (Corbet \& Peele 2001) which may be an orbit
period. Recent SWIFT/BAT observations (Corbet et al. 2007) and
RXTE/ASM data (Wang 2009) found a modulation of $\sim 19.12$ days
consistent with twice the 9.6-day period.

The nature of the compact object in 4U 2206+54 has been in dispute
for a long time (Negueruela \& Reig 2001; Corbet \& Peele 2001).
Recent reports on the detection of cyclotron resonant absorption
line at $\sim 30$ keV and 60 keV suggested a magnetized neutron
star with a magnetic field of $\sim 3.3\times 10^{12}$ G by
INTEGRAL observations (Wang 2009; Blay et al. 2005). The reports
of possible 5500 s pulsations in light curves of 4U 2206+54 from
RXTE, INTEGRAL and Suzaku observations suggested that it would be
a X-ray pulsar (Reig et al. 2009; Wang 2009; Finger et al. 2009).

In this work, we will report a long duration hard X-ray flare in
4U 2206+54 discovered by INTEGRAL/IBIS and RXTE/ASM observations.
We first introduce the observations of INTEGRAL/IBIS and RXTE/ASM
telescopes in \S 2. The main features of the hard X-ray flare in
4U 2206+54 are described in \S 3. The summary and discussions on
the possible origins of this hard X-ray flare will be delineated
in \S 4.

\section{Observations}

\subsection{INTEGRAL/IBIS}

The hard X-ray source 4U 2206+54 was observed during the INTEGRAL
pointed observations of the Cassiopeia region around December
2005. The hard X-ray flare were captured by the low-energy
detector (called ISGRI) of the Imager (IBIS, Lebrun et al. 2003)
aboard INTEGRAL. The IBIS-ISGRI scientific data analysis was
carried out using the Off-line Scientific Analysis (OSA) software
version 7.0 (Goldwurn et al. 2003) provided by the INTEGRAL
Science Data Center (ISDC). Individual pointings in each satellite
revolution (3 days) processed with OSA 7.0 were mosaicked to
create sky images for source detections. We have used the 20 -- 40
keV band for source detection and to quote fluxes (Table 1). For
Rev 387, 4U 2206+54 has a detection significance level of $\sim
45\sigma$ with an average IBIS/ISGRI count rate up to $\sim 11.9$
cts/s in the energy band of 20 -- 40 keV. 4U 2206+54 appeared in
the outburst state around Dec 15 2005 in the hard X-ray bands.

We derived the hard X-ray light curve of 4U 2206+54 in the band of
20 -- 40 keV from 2005 Dec 11 to Dec 19 (see Fig. 1). 4U 2206+54
was unfortunately outside the field of view (FOV) of IBIS
sometimes, and the data were also screened for solar-flare events
and erratic count fluctuations due to passages through the Earth's
radiation belts, so several data gaps appeared in the light curve
of 4U 2206+54 (Fig. 1). From the available IBIS data, a strong
hard X-ray outburst was still detected around Dec 15 -- 16, 2005,
which lasted $\sim 10^5$ s.


\subsection{RXTE/ASM}

RXTE had no pointed observations on the source 4U 2206+54 during
December 2005. Fortunately the source was regularly monitored by
the All Sky Monitor (ASM) onboard RXTE. The ASM consists of three
Scanning Shadow Cameras (SSCs) mounted on a rotating drive
assembly. Each Camera has a field of view of $6^\circ \times
90^\circ$. The assembly "dwells" at a fixed position for $\sim$ 90
s, followed by a rotation of 6$^\circ$. Each camera has a position
sensitive proportional counter, and the data is analyzed to give
not only the total source intensity in the 1.5 -- 12 keV band, but
also the intensity in each of 3 energy bands: 1.5 -- 3 keV, 3 -- 5
keV and 5 -- 12 keV. The ASM data are available in two forms:
count rates from individual 90-s dwells, and one-day average for a
source.


From the archival dataset
\footnote{http://xte.mit.edu/asmlc/ASM.html} provided by the
surveys of the All-Sky Monitor (ASM) on board RXTE, we obtained
the dwell-by-dwell light curve (1.5 -- 12 keV) of 4U 2206+54
during the hard X-ray outburst (Fig. 1). ASM data filled up the
data gap before the hard X-ray burst obtained from the IBIS data.
Therefore, the complete observed features of the super-long
duration hard X-ray outburst in 4U 2206+54 were derived by using
both ASM and IBIS observations.

\begin{table*}

\caption{INTEGRAL/IBIS observations of the field around 4U 2206+54
from 2005 Dec to 2006 Jan. The time intervals of observations in
the revolution number and the corresponding dates, the corrected
on-source exposure time are listed. Mean count rate and the
detection significance level value in the energy range of 20 -- 40
keV were also shown. }

\begin{center}
\scriptsize
\begin{tabular}{l c c c l}

\hline \hline Rev. Num. & Date  & On-source time (ks) & Mean count rate s $^{-1}$ & Detection level \\
\hline
384 & 2005 Dec 05 -- 07 & 118  &  $2.86\pm 0.24$ & 15$\sigma$ \\
385 & 2005 Dec 08 -- 10 & 132 & $2.01\pm 0.20$ & 11$\sigma$\\
386 & 2005 Dec 11 -- 13 & 121 & 1.66$\pm 0.19$ & 10$\sigma$ \\
387 & 2005 Dec 14 -- 16 & 98 & $11.9\pm 0.25$ & 45$\sigma$ \\
388 & 2005 Dec 17 -- 19 & 103 & $2.38\pm 0.21$ & 14$\sigma$ \\
389 & 2005 Dec 20 -- 22 & 110 & $1.38\pm 0.30$ & 6$\sigma$ \\
390 & 2005 Dec 23 -- 25 & 102 & 0.66$\pm 0.25$ & 4.9$\sigma$ \\
391 & 2005 Dec 26 -- 28 & 113 & 1.78$\pm 0.21$ & 9$\sigma$ \\
392 & 2005 Dec 29 -- 31 & 110 & $2.58\pm 0.22$ & $14\sigma$ \\
393 & 2006 Jan 01 -- 03 & 118 & $0.77\pm 0.21$ & $5.5\sigma$\\
394 & 2006 Jan 04 -- 06 & 115 & $1.68\pm 0.20$ & 10$\sigma$\\
395  & 2006 Jan 07 -- 09 & 124 & $1.61\pm 0.21$ & 9$\sigma$\\
\hline
\end{tabular}
\end{center}

\end{table*}

\section{Results}

The hard X-ray outburst occurred at UT 2005 Dec 15.1085 and lasted
over 2 days. This peculiar flare showed a double-flare feature in
the light curve. We will present the properties of the first and
second flares separately in the following.

\subsection{The first flare}

The first flare was only captured by the ASM data and had a peak
flux of $\sim$ 100 mCrab and decayed to $\sim$ 20 mCrab. This
flare appeared to have a sharp rise peak and then the flux decayed
with a long time of $\sim 15$ hours. The peak flux in the energy
range of 1.5 --12 keV is $(3.3\pm 0.9)\times 10^{-9}$ erg
cm$^{-2}$ s$^{-1}$, and a total fluence $\sim 5.3\times 10^{-5}$
erg cm$^{-2}$. For a distance of 2.6 kpc (Blay et al. 2006), the
peak luminosity at the first flare reached $\sim 3.6\times
10^{36}$ erg s$^{-1}$, and total energy release during the first
flare is $\sim 6\times 10^{40}$ ergs in the energy range of
1.5--12 keV. The hardness ratio analysis from ASM data implied a
hard spectrum in the first flare: this X-ray flare cannot be
significantly detected below 5 keV (Fig. 2).

\begin{figure}
\includegraphics[angle=0,width=9cm]{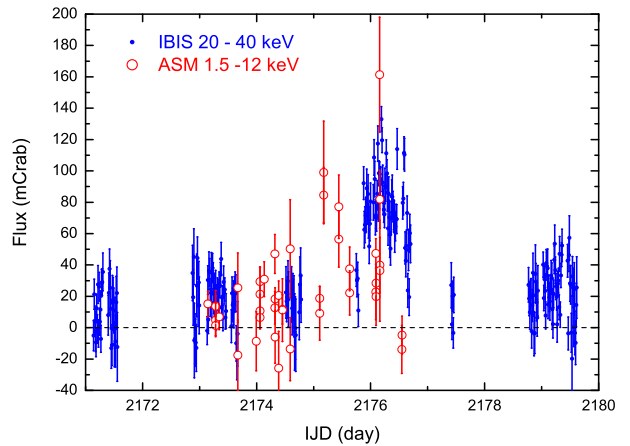}
\caption{Hard X-ray lightcurves of 4U 2206+54 from IJD 2171 to
2180 (IJD=MJD-51544) in the energy ranges of 1.5 -12 keV from the
RXTE/ASM data and 20--40 keV observed by INTEGRAL/IBIS. The flux
unit of 1 Crab corresponds to 2.2$\times 10^{-8}$ erg cm$^{-2}$
s$^{-1}$ in the band of 1.5 --12 keV and 7.6$\times 10^{-9}$ erg
cm$^{-2}$ s$^{-1}$ in the 20-- 40 keV band.  }
\end{figure}

\subsection{The second flare}

The second flare appeared about $\sim$ 15 hours after the first
one. It was observed by both ASM and IBIS. ASM only obtained a few
data points for the second flare but IBIS showed the detailed
variability information (Fig. 3).  In the energy range of 20 -- 40
keV, the second flare lasted more than 8$\times 10^4$ second.
Multiple mini-peak feature appeared in the second flare. The power
spectrum analysis showed a significant modulation period at $\sim
5550\pm 50$ s (Fig. 3), which confirms the pulsation period of the
neutron star in 4U 2206+54 (Reig et al. 2009; Wang 2009). The
folded light curve is shown in Fig. 4. The pulse fraction defined
as the ratio of the pulse maximum minus the minimum to the maximum
is estimated to be $\sim 30\%$ during the flare. Reig et al.
(2009) found a pulse fraction of $\sim 50\%$ with RXTE and
INTEGRAL data; and Wang (2009) obtained a pulse fraction of $\sim
80\%$. The pulse fraction during the outburst appeared to be lower
than those in the other states.

\begin{figure}
\includegraphics[angle=0,width=7.5cm]{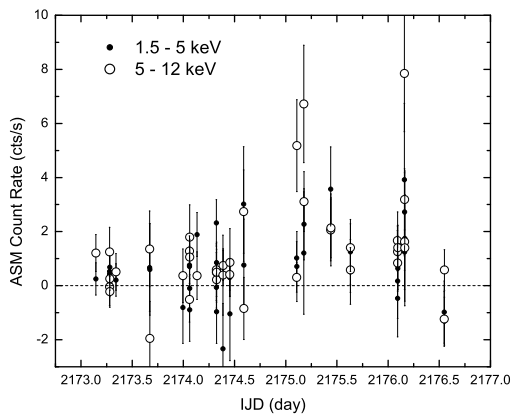}
\caption{The ASM lightcurves of the X-ray binary 4U 2206+54 in two
energy bands: 1.5--5 keV; 5 --12 keV. The X-ray flare cannot be
significantly detected below 5 keV. }
\end{figure}

We obtained the spectrum for the second flare from 20 -- 300 keV
from IBIS observations (see Fig. 5). This spectrum can be fitted
well by two different models: a thermal bremsstrahlung model with
$kT \sim 23.4\pm 1.3$ keV (reduced $\chi^2\sim 0.81$, 9 d.o.f.);
and a broken power-law model with the photon indexes $\Gamma_1
\sim 2.3\pm 0.2,\ \Gamma_2 \sim 3.3\pm 0.3$, and the break energy
is $E_b \sim 31.2\pm 1.7$ keV (reduced $\chi^2\sim 0.77$, 7
d.o.f.). No cyclotron absorption lines are found in the spectrum
of the different fits.

The average flux during the second flare from 20 -- 150 keV energy
band is $\sim (1.2\pm 0.1)\times 10^{-9}$ erg cm$^{-2}$ s$^{-1}$,
corresponding to a mean luminosity of 1.3$\times 10^{36}$ erg
s$^{-1}$ in the range of 20 --150 keV assuming a distance of 2.6
kpc. Then the total released energy of the second flare with the
duration of $\sim 10^5$ sec is about 10$^{41}$ ergs.

\begin{figure}
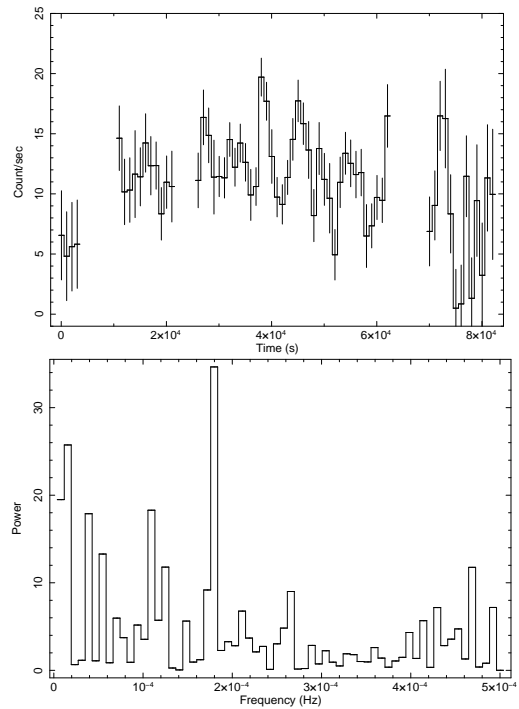

\includegraphics[angle=-90,width=6.5cm]{4u2206lc.ps}
\includegraphics[angle=-90,width=6.8cm]{4u2206powsp.ps}
\caption{{\bf Top} The light curve of the second flare of the hard
X-ray outburst by INTEGRAL/IBIS.  {\bf Bottom} The power spectrum
of the INTEGRAL/IBIS light curve of 4U 2206+54 during the second
flare. A significant period at $5550\pm 50$ s is detected in the
light curve.}
\end{figure}

\begin{figure}
\includegraphics[angle=-90,width=6.5cm]{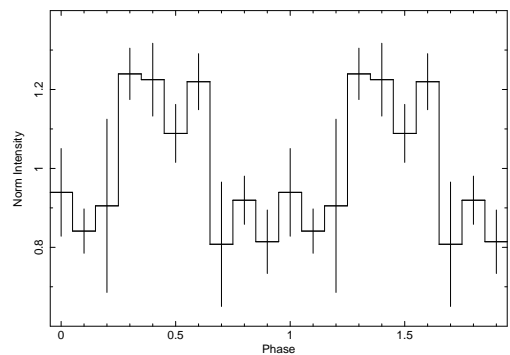}
\caption{The IBIS/ISGRI background subtracted light curve (20 --
40 keV) of 4U 2206+54 during the second flare folded at a
pulsation period (5550 s). The pulse profile is repeated once for
clarity.}
\end{figure}

\begin{figure}
\includegraphics[angle=-90,width=6.5cm]{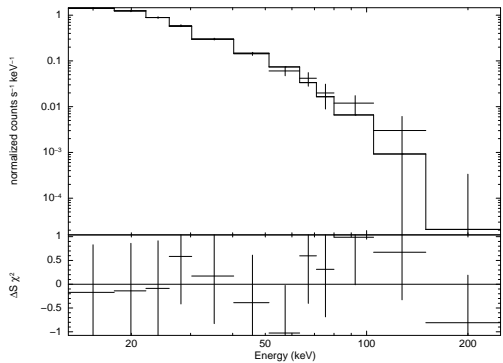}
\caption{The average spectrum of 4U 2206+54 from the INTEGRAL/IBIS
observations during the second flare of the X-ray burst. The
spectrum was fitted by a thermal bremsstrahlung model with $kT\sim
23.2$ keV. }
\end{figure}

\section{Summary and discussion}

In this paper we reported the INTEGRAL/IBIS and RXTE/ASM
observations of a 2-day long-duration hard X-ray outburst in 4U
2206+54 which harbors a highly magnetized neutron star (Wang
2009). This outburst is peculiar in the following ways: (1) a
double-flare feature: the second flare occurred about 15 hours
after the first one; the first flare had a temporal profile of
fast rise and exponential like decay, the second one had a
modulation period of 5550 s which is the pulsation period of the
neutron star; (2) the flare had a hard spectrum: the first flare
was only seen above 5 keV from RXTE/ASM data; for the second
flare, the 18 -- 250 keV spectrum was fitted by a thermal
bremsstrahlung model with $kT\sim 23.4$ keV or a broken power-law
model with the photon indexes $\Gamma_1 \sim 2.3,\ \Gamma_2 \sim
3.3$, and break energy $E_b \sim 31$ keV. No cyclotron absorption
line during the outburst was detected. The long-duration flare has
a mean X-ray luminosity of $\sim 1.3\times 10^{36}$ erg s$^{-1}$
and a total released energy higher than $10^{41}$ ergs in the
range of 20 -- 100 keV.

The sudden strong X-ray flares have been detected in some other
high mass X-ray binaries (e.g., GX 301-2, see Haberl 1991, Leahy
1991, Koh et al. 1997; XTE J0052-723 Laycock et al 2003). In these
systems, the X-ray flares occur always before periastron passages
of the neutron star. Therefore, the flares show the recurrence of
orbital periodicity, and have the following characteristics: (1)
duration of $\sim 0.1$ orbital phase, roughly several days (Leahy
2002; Laycock et al 2003); (2) strong photoelectric absorption,
which makes the flares events absent at low energies ($<5$ keV,
see Leahy 2002); (3) the spectrum can be described by a cut-off
power-law model (Laycock et al 2003). Recently, the X-ray outburst
of a Be/X-ray pulsars A 0535+26 was detected with a double-peak
feature (Caballero et al. 2010). This outburst also showed the
orbital modulations with a duration of more than ten days (orbital
period $\sim 111$ days, Finger et al. 2006). In addition the
cyclotron absorption features were detected during the outburst,
suggesting that the transient circumstellar-disk accretion near
the periastron powered the outburst of A 0535+26 (Caballero et al.
2010).



Since these X-ray flares occurred with orbital periodicity, we can
check this possible modulation in X-ray light curve of 4U 2206+54.
If we assume the detected X-ray flare occur around the periastron
passage of the neutron star, then the X-ray flares would be
recurrent with the recurrence time of $\sim 19$ day. In Table 1,
we presented the hard X-ray light curves (20 -- 40 keV) of 4U
2206+54 observed by IBIS-ISGRI covering two orbital periods, but
only one flare was detected and no orbital modulation was found.
Then the X-ray flare in 4U 2206+54 was different from the
outbursts near the periastrons in other sources because the
detected hard X-ray flare would not be repeated just due to
orbital modulation. Thus the hard X-ray flare in 4U 2206+54 may
have other origins.

Recently, some soft gamma-ray time-structured bursts of durations
from several hours to about 1 day have been detected by
INTEGRAL/IBIS observations in some supergiant high mass X-ray
binaries which were called supergiant fast X-ray transients
(SFXTs, see Sidoli et al. 2005; Negueruela et al. 2006; Sguera et
al. 2006; 2008). The SFXTs also belong to the wind-fed systems.
The physical origin of the fast outbursts displayed in SFXTs is
still unknown. It was suggested that the presence of dense clumps
in the wind of OB supergiant companions produces the accretion
outbursts in SFXTs (in't Zand 2005; Walter \& Zurita Heras 2007).
It is believed that SFXTs should contain sporadically accreting
neutron stars, and they would be similar to the system of 4U
2206+54. Some of SFXTs have been found to contain a neutron star
with pulsation period from ten to a few hundred seconds (Sguera et
al. 2007; Karasev et al. 2008), and orbital period from several
days to 30 days (Bird et al. 2009; Jain et al. 2009). 4U 2206+54
has a similar orbital period but a much slower pulsation period.
The detected flare in 4U 2206+54 also has the peak X-ray
luminosity similar to those of SFXTs. Then the reported outbursts
with X-ray luminosity higher than 10$^{36}$ erg s$^{-1}$ could
just be a bright flare in 4U 2206+54 due to a suddenly enhanced
wind accretion induced by an unknown process, which might be
similar to the theoretical mechanism proposed in SFXTs. It is
possible that the accretion materials as clumps or conglomeration
from stellar winds hit the polar cap region of the highly magnetic
neutron star to produce the hard X-ray outburst in 4U 2206+54.

There still exist the different effects on accretion by the wind
of a main sequence star in 4U 2206+54 and supergiant winds in
SFXTs. More detailed study is clearly needed. It should be noted
that a small part of the unidentified sources show the fast X-ray
bursts and are candidate SFXTs though the optical counterpart has
not yet been identified as the early-type supergiant star (see
Negueruela et al. 2006; Sguera et al. 2006). Moreover, many
early-type stars are characterized by highly structured and
variable massive winds (Prinja et al. 2005). It is expected that
this hard X-ray flare in 4U 2206+54 could be a {\it missing} link
between these systems, and may help us to understand the nature of
SFXTs and the flare of 4U 2206+54.

\section*{Acknowledgments}
The author is grateful to the referee for the fruitful comments
and Prof. C.K. Chou for useful discussions. This work was
supported by the National Natural Science Foundation of China
under grants 10803009 and 10833003.




\end{document}